# THE PROCEDURE OF EXAMINATION OF THE STABILITY OF BOUNDARY BETWEEN PLASMA AND THE MAGNETIC FIELD IN ELECTRONEUTRAL APPROACH

V.V. Lyahov, V.M. Neshchadim

Institute of Ionosphere, Kamenskoe plato, 050020, Almaty, Kazakhstan[(1,2)]

## 1. Introduction

Study of the structure of equilibrium boundary layers between both cold and thermal plasma and magnetic field has long history and begins with papers of Chapman and Ferraro [1]. The authors naturally deal with the effect of polarization of the boundary layers that originates in magnetized plasma due to different mass values and consequently Larmor radiuses of electronic and ionic components. Many solutions, for example [2], were found in quasi-neutrality approximation. Of course, the electric field originating in such approximate solutions calls for further investigations and full solution of the problem. Other solutions are found in the electroneutral class of solutions [3,4], which is achieved by certain constraining of problem parameters. Some researchers, for example [5], postulate the integral neutrality of transition layers. Finally, in a number of papers (for example [6, 7]) a conclusion was made that plasma of transition layers is generally polarized.

In the study of stability of 'plasma – magnetic field' boundary, preliminary strict investigation of the equilibrium of this boundary is usually neglected. As equilibrium solution in the textbook [8], for example, is a priori taken the Maxwell distribution function with non-uniform values of density and temperature. In particular, the case where the characteristic size of inhomogenuity of boundary much more exceeds the Larmor radiuses of components of plasma is investigated. Nevertheless, general boundary problem needs further investigation. Such statement of the problem leaves self-consistency of electromagnetic fields as well as degree of adequacy of the proposed models to the real boundary and its stability outside the limits of investigation. Thus, in the class of self-consistent solutions plasma with Maxwell velocity on velocities is unable to generate an equilibrium boundary with containing magnetic field.

## 2. Statement of the Problem

Thus, despite success in study of a large number of instabilities of magnetic plasma, the vital question is development of the method of study of equilibrium and stability of boundary layers between plasma and magnetic field, which should be viewed as self-consistent and be of arbitrary thickness (the characteristic thickness is determined by plasma and magnetic field parameters). Before taking on this physical problem, it is necessary to create necessary pre-requisites: 1) to solve kinetic equation with a self-consistent electromagnetic field for perturbation of distribution function, 2) to calculate on the basis of this solution tensor of dielectric permeability necessary for a constitutive equation, 3) to derive a dispersion equation from the Maxwell system of equations closed with the derived constitutive equation and develop a technique to analyze it.

---

[(1,2)] mail: v_lyahov@rambler.ru, ne_dim@bk.ru

Let plasma concentrate in the inferior half-space (refer to Fig. 1) and containing it field concentrate in the upper half-space. On the boundary of these fields in plane *xoy* the self-consistent transition layer generates.

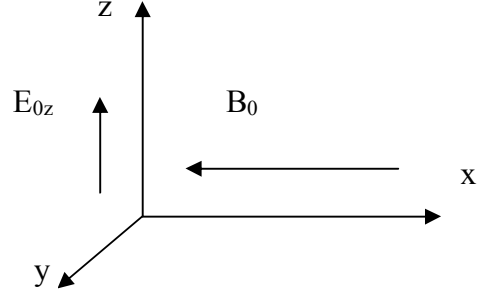

Fig. 1

The problem is one-dimensional and all quantities depend on variable *z*. The investigated medium is described by the system from the kinetic equation

$$\frac{\partial f_\alpha}{\partial t} + \vec{v}\frac{\partial f_\alpha}{\partial \vec{r}} + e_\alpha \{\vec{E} + [\vec{v}\vec{B}]\}\frac{\partial f_\alpha}{\partial \vec{P}_\alpha} = 0, \qquad (1)$$

and Maxwell equations with a self-consistent electromagnetic field (no external sources) (equation 1)

$$rot\vec{B} = \frac{1}{c^2}\frac{\partial \vec{E}}{\partial t} + \frac{1}{\varepsilon_0 c^2}\vec{j}, \quad div\vec{B} = 0, \qquad (2)$$

$$rot\vec{E} = -\frac{\partial \vec{B}}{\partial t}, \quad div\vec{E} = \frac{\rho}{\varepsilon_0},$$

where где

$$\rho = \sum_\alpha e_\alpha \int \delta f_\alpha d\vec{P},$$

$$\vec{j} = \sum_\alpha e_\alpha \int \vec{v}\delta f_\alpha d\vec{P}.$$

The problem (1), (2) is solved by methods of perturbation theory

$$f_\alpha(\vec{P},\vec{r},z,t) = f_{0\alpha}(\vec{P}) + \delta f_\alpha(\vec{P},\vec{r},z,t);$$

$$\vec{E}(\vec{r},z,t) = \vec{E}_0(z) + \delta\vec{E}(\vec{r},z,t); \qquad (3)$$

$$\vec{B}(\vec{r},z,t) = \vec{B}_0(z) + \delta\vec{B}(\vec{r},z,t).$$

Plasma is considered to be weakly non-equilibrium $\delta f_\alpha(\vec{P},\vec{r},z,t) < f_{0\alpha}(\vec{P})$.

### 3. Solution of kinetic equation

#### 3.1. Study of equilibrium of a transition layer

Equilibrium distribution functions is devised as function of motion integrals $f_{0\alpha}(\vec{P}) = f_{0\alpha}(W, P_y, P_x)$ where the total energy and generalized momentum take the form of:

$$W = \frac{1}{2} m_\alpha (v_x^2 + v_y^2 + v_z^2) + e_\alpha \phi(z),$$
$$P_y = m_\alpha v_y + e_\alpha A_y(z), \qquad (4)$$
$$P_x = m_\alpha v_x.$$

Here, $\phi(z), A_y(z)$ - electrical and magnetic potentials ($\vec{E}_0 = -grad\phi, \vec{B}_0 = rot\vec{A}$).

Selection of equilibrium distribution function in the form of [3]

$$f_{0\alpha}(W, P_y) = \left(\frac{m_\alpha}{2\pi\theta_{\alpha z}}\right)^{\frac{3}{2}} n_{0\alpha}(1+a_\alpha) \exp\{-\frac{W_\alpha}{\theta_{\alpha z}} - \frac{a_\alpha P_{y\alpha}}{2m_\alpha \theta_{\alpha z}}\}, \qquad (5)$$

where degree of anisotropy $a_\alpha$ is determined by formula

$$a_\alpha = (\frac{\theta_{\alpha z}}{\theta_{\alpha y}} - 1), \qquad (6)$$

and motion integrals $W_\alpha$ and $P_{y\alpha}$ are determined by formula (4) allows derivation of equations for potentials of a self-consistent electromagnetic field:

$$\eta \frac{d^2\psi}{d\xi^2} = \exp(\frac{\psi}{\beta} - \frac{\alpha_e}{1+\alpha_e}\frac{a^2}{2\beta\mu}) - \exp(-\psi - \frac{\alpha_i}{1+\alpha_i}\frac{a^2}{2}), \qquad (7)$$

$$\frac{d^2 a}{d\xi^2} = \frac{\alpha_e}{1+\alpha_e}\frac{a}{\mu} \exp(\frac{\psi}{\beta} - \frac{\alpha_e}{1+\alpha_e}\frac{a^2}{2\beta\mu}) + \frac{\alpha_i}{1+\alpha_i} a \exp(-\psi - \frac{\alpha_i}{1+\alpha_i}\frac{a^2}{2}). \qquad (8)$$

Boundary conditions are:

$$\psi(-\infty) = 0, a(-\infty) = 0, E(-\infty) = -\psi'(-\infty) = 0, B(-\infty) = a'(-\infty) = 0. \qquad (9)$$

The following non-dimensional quantities are introduced:

$$\mu = \frac{m}{M}, \beta = \frac{\theta_{xe}}{\theta_{xi}}, \eta = \frac{\theta_{xi}}{Mc^2}, \psi = \frac{e}{\theta_{xi}}\phi, a = \frac{e}{c}\frac{1}{\sqrt{\theta_{xi}M}} A_y, \xi^2 = \frac{4\pi e^2 n_0}{Mc^2} x^2. \qquad (10)$$

The numerical solution of problem (7), (8), (9) will give trivial null. It is necessary first to find asymptotics of equations (7), (8) in the field of $a \to 0, \psi \to 0$. Leaving small terms of the order of $\psi$ and $a^2$, we may find asymptotics of the system:

$$\frac{d^2\psi}{d\xi^2} = \frac{1+\beta}{\beta\eta}\psi + \frac{1}{2\eta}(\frac{\alpha_i}{1+\alpha_i} - \frac{1}{\beta\mu}\frac{\alpha_e}{1+\alpha_e})a^2,$$

$$\frac{d^2 a}{d\xi^2} = (\frac{1}{\mu}\frac{a_e}{1+\alpha_e} + \frac{\alpha_i}{1+\alpha_i})a. \qquad (11)$$

The solution of this system is:

$$a = c_2 \exp(\sqrt{b}\xi), B = a' = c_2 \sqrt{b} \exp(\sqrt{b}\xi),$$

$$\psi = \frac{c_2 d}{2\sqrt{s}(\sqrt{b}-\sqrt{s})} \exp(\sqrt{b}\xi) + c_3 \exp(\sqrt{s}\xi), \qquad (12)$$

$$E = -\psi' = -\frac{c_2 d \sqrt{b}}{2\sqrt{s}(\sqrt{b}-\sqrt{s})} \exp(\sqrt{b}\xi) - c_3 \sqrt{s} \exp(\sqrt{s}\xi).$$

At the numerical calculations we should begin from the asymptotic solution (12) in some point $\xi$. Here, designations are introduced:

$$s = \frac{1+\beta}{\beta\eta}, d = \frac{1}{2\eta}(\frac{\alpha_i}{1+\alpha_i} - \frac{1}{\beta\mu}\frac{\alpha_e}{1+\alpha_e}), b = (\frac{1}{\mu}\frac{a_e}{1+\alpha_e} + \frac{\alpha_i}{1+\alpha_i}). \quad (13)$$

Coefficients $c_2$ and $c_3$ can be arbitrary; in particular we can set:

$c_2 = c_3 = 1$.

In general, plasma of a stationary boundary layer is polarized, and this electrical polarization field should be taken into account in the study of unstable stability of boundary layers.

### 3.2. Determination of non-equilibrium component to distribution function $\delta f_\alpha$

Knowing now an equilibrium distribution function (5) and the equilibrium electrical and magnetic fields defined by the solution of equations (7), (8), (9), we can derive equation for determination of a non-equilibrium component to distribution function $\delta f_\alpha$. Substituting expansions (3) in equation (1) and taking into account the smallness of non-equilibrium components, we will obtain a linear kinetic equation for $\delta f_\alpha$:

$$\frac{\partial \delta f_\alpha}{\partial t} + \vec{v}\frac{\partial \delta f_\alpha}{\partial \vec{r}} + e_\alpha \{\vec{E}_0 + [\vec{v}\vec{B}_0]\}\frac{\partial \delta f_\alpha}{\partial \vec{P}_\alpha} = -e_\alpha \{\delta\vec{E} + [\vec{v}\delta\vec{B}]\}\frac{\partial f_{0\alpha}}{\partial \vec{P}_\alpha}. \quad (14)$$

Owing to the linearity of equation (14) and field equations (2), dependence of all perturbed quantities on time and co-ordinates will be presented as:

$$\delta f_\alpha(\vec{P},\vec{r},z,t) = \delta f_\alpha(\vec{P},\omega,\vec{k},z)\exp(-i\omega t + ik_x x + ik_y y),$$

$$\delta\vec{E}(\vec{r},z,t) = \delta\vec{E}(\omega,\vec{k},z)\exp(-i\omega t + ik_x x + ik_y y), \quad (15)$$

$$\delta\vec{B}(\vec{r},z,t) = \delta\vec{B}(\omega,\vec{k},z)\exp(-i\omega t + ik_x x + ik_y y).$$

After substituting expansions (15) in equation (14), we will obtain equation for Fourier amplitudes:

$$-i(\omega - \vec{k}\vec{v})\delta f_\alpha + v_z\frac{\partial \delta f_\alpha}{\partial z} + e_\alpha\{\vec{E}_0 + [\vec{v}\vec{B}_0]\}\frac{\partial \delta f_\alpha}{\partial \vec{P}_\alpha} + e_\alpha \delta\vec{E}\frac{\partial f_{0\alpha}}{\partial \vec{P}_\alpha} = 0. \quad (16)$$

In study of perturbation of distribution function $\delta f_\alpha$ we will confine ourselves to investigation of its evolution in momentum space. In the configuration space function $\delta f_\alpha(\vec{P},\vec{r},z,t)$ depends on co-ordinate z, only as from a parameter (see (15)) as the viewed perturbations do not spread along Z-axis. Dependence of transition layer characteristics on z has been studied comprehensively at equilibrium solution development, and this dependence graded in the study of stability of boundary layer. Therefore, to simplify the situation we will neglect the second term in the last equation. In the momentum space we will pass, as usually, to a cylindrical coordinates system: $P_\perp, \varphi, P_z$ ($P_y = P_\perp \cos\varphi, P_z = P_\perp \sin\varphi$), then last equation will take the form of:

$$\frac{\partial \delta f_\alpha}{\partial \varphi} - \frac{e_\alpha E_{0z}\sin\varphi}{\Omega_\alpha(z)}\frac{\partial \delta f_\alpha}{\partial P_{\alpha\perp}} = -\frac{i(\omega-\vec{k}\vec{v})}{\Omega_\alpha(z)}\delta f_\alpha + \frac{e_\alpha}{\Omega_\alpha(z)}\delta\vec{E}\frac{\partial f_{0\alpha}}{\partial \vec{P}_\alpha}. \quad (17)$$

Here, the formula was used

$$E_{0\perp} = E_{0z} \sin \varphi \qquad (18)$$

and the Larmor frequency was introduced

$$\Omega_\alpha(z) = \frac{e_\alpha B_0(z)}{m_\alpha}.$$

Solution of equation (17) in case plasma polarization is disregarded is known as:

$$\delta f_\alpha = \frac{e_\alpha}{\Omega_\alpha(z)} \int_\infty^\varphi (\delta \vec{E} \frac{\partial f_{0\alpha}}{\partial \vec{P}_\alpha})_{\varphi'} \exp[\frac{i}{\Omega_\alpha(z)} \int_\varphi^{\varphi'}(\omega - \vec{k}\vec{v})_{\varphi''} d\varphi''] d\varphi' . \qquad (19)$$

That is, application of perturbation technique (3) to kinetic equation with self-consistent field (1), (2) made it possible to consistently defined equilibrium distribution function (5) and non-equilibrium component to it (19). The theory of boundary equilibrium between plasma and magnetic field retaining is developed on the basis of equilibrium distribution function. Non-equilibrium distribution function (19) can be taken as a basis for analysis of stability of this boundary.

### 4. Study of transition layer stability

#### 4.1. Evaluation of a tensor of an inductivity

Substituting the determined non-equilibrium correction for distribution function $\delta f_\alpha$ (19) in the equation of current density induced in plasma:

$$\vec{j} = \sum_\alpha e_\alpha \int \vec{v} \delta f_\alpha d\vec{P},$$

and using the constitutive equation of medium in the form of:

$$j_i = \sigma_{ij} \delta E_j, \qquad (20)$$

it is possible to compute tensor of conductivity $\sigma_{ij}$, and then, using formula:

$$\varepsilon_{ij} = \delta_{ij} + \frac{i}{\varepsilon_0 \omega} \sigma_{ij}, \qquad (21)$$

it is possible to compute tensor of dielectric permeability.

We will make calculation in the Cartesian coordinate system:

$$\varepsilon_{ij} = \delta_{ij} + \frac{i}{\varepsilon_0 \omega} \sigma_{ij} = \delta_{ij} + \frac{i}{\varepsilon_0 \omega} \sum_\alpha \frac{e_\alpha^2}{m_\alpha \Omega_\alpha(z)} \int d\vec{v} v_i \int_\infty^\varphi d\varphi' (\frac{\partial f_{0\alpha}}{\partial v_j})_{\varphi'} \exp[\frac{i}{\Omega_\alpha(z)} \int_\varphi^{\varphi'} d\varphi''(\omega - \vec{k}\vec{v})_{\varphi''}] \quad (22)$$

We shall find derivatives with respect velocities of the equilibrium distribution function (5) that are necessary for calculation of tensor $\varepsilon_{ij}$ (22):

$$\frac{\partial f_{0\alpha}}{\partial v_x} = \left(\frac{m_\alpha}{2\pi\theta_\alpha}\right)^{\frac{3}{2}} n_{0\alpha}(1+\alpha_\alpha)^{\frac{1}{2}} \left[-\frac{m_\alpha v_x}{\theta_\alpha}\right] \exp\left[-\frac{W_\alpha}{\theta_\alpha} - \frac{\alpha_\alpha m_\alpha V_{y\alpha}^2}{2\theta_\alpha}\right],$$

$$\frac{\partial f_{0\alpha}}{\partial v_y} = \left(\frac{m_\alpha}{2\pi\theta_\alpha}\right)^{\frac{3}{2}} n_{0\alpha}(1+\alpha_\alpha)^{\frac{1}{2}} \left[-\frac{m_\alpha v_y}{\theta_\alpha} - \frac{\alpha_\alpha m_\alpha (v_y + \frac{e_\alpha}{m_\alpha} A_y(z))}{2\theta_\alpha}\right] \exp\left[-\frac{W_\alpha}{\theta_\alpha} - \frac{\alpha_\alpha m_\alpha V_{y\alpha}^2}{2\theta_\alpha}\right],$$

$$\frac{\partial f_{0\alpha}}{\partial v_z} = \left(\frac{m_\alpha}{2\pi\theta_\alpha}\right)^{\frac{3}{2}} n_{0\alpha}(1+\alpha_\alpha)^{\frac{1}{2}} \left[-\frac{m_\alpha v_z}{\theta_\alpha}\right]\exp\left[-\frac{W_\alpha}{\theta_\alpha} - \frac{\alpha_\alpha m_\alpha V_{y\alpha}^2}{2\theta_\alpha}\right]. \quad (23)$$

We will first perform the angle integration. If the last integral in formula (22) is taken, it will rearrange in the form of:

$$\varepsilon_{ij} = \delta_{ij} + \frac{i}{\varepsilon_0 \omega}\sum_\alpha \frac{e_\alpha^2}{m_\alpha \Omega_\alpha(z)}\int d\vec{v}\exp[-i\frac{\omega - k_x v_x}{\Omega_\alpha(z)}\varphi]\exp[ib_\alpha \sin\varphi]v_i\int_\infty^\varphi (\frac{\partial f_{0\alpha}}{\partial v_j})_{\varphi'} \cdot$$
$$\exp[i\frac{\omega - k_x v_x}{\Omega_\alpha(z)}\varphi']\exp[-ib_\alpha \sin\varphi']d\varphi', \quad (24)$$

where

$$b_\alpha = \frac{k_\perp v_\perp}{\Omega_\alpha(z)}. \quad (25)$$

We will describe in detail the calculation of one of the components of tensor of dielectric permeability, for example, component $\varepsilon_{xx}$.

Substituting the first relation (23) in (24), it can be obtained:

$$\varepsilon_{xx} = 1 + A\exp[-\frac{m_\alpha}{2\theta_\alpha}(v_x^2 + v_\perp^2)]\exp[-\frac{m_\alpha}{2\theta_\alpha}(\frac{\alpha_\alpha e_\alpha^2 A_y^2(z)}{m_\alpha^2})] \cdot$$
$$\int_\infty^\varphi d\varphi' \exp[-ib_\alpha \sin\varphi']\exp[i\frac{\omega - k_x v_x}{\Omega_\alpha(z)}\varphi'] \cdot \quad (26)$$
$$\exp[-\frac{m_\alpha}{2\theta_\alpha}(\alpha_\alpha v_\perp^2 \cos^2\varphi' + 2\alpha_\alpha \frac{e_\alpha}{m_\alpha}A_y(z)v_\perp \cos\varphi')].$$

Here

$$A = \frac{i}{\varepsilon_0 \omega}\sum_\alpha \frac{e_\alpha^2}{m_\alpha \Omega_\alpha(z)}\int d\vec{v}\exp[-i\frac{\omega - k_x v_x}{\Omega_\alpha(z)}\varphi]\exp[ib_\alpha \sin\varphi](-\frac{m_\alpha v_x^2}{\theta_\alpha}) \cdot$$
$$\left(\frac{m_\alpha}{2\pi\theta_\alpha}\right)^{\frac{3}{2}} n_{0\alpha}(1+\alpha_\alpha)^{\frac{1}{2}}. \quad (27)$$

Using of the expansion:

$$\exp[-ib_\alpha \sin\varphi'] = \sum_{n=-\infty}^{\infty} J_n(b_\alpha)\exp[-in\varphi'], \quad (28)$$

where $J_n$ - Bessel function of $n$-order,

we will derive from (26):

$$\varepsilon_{xx} = 1 + A\exp[-\frac{m_\alpha}{2\theta_\alpha}(v_x^2 + v_\perp^2)]\exp[-\frac{m_\alpha}{2\theta_\alpha}(\frac{\alpha_\alpha e_\alpha^2 A_y^2(z)}{m_\alpha^2})]\cdot$$

$$\sum_n J_n(b_\alpha)\int_\infty^\varphi d\varphi' \exp[i(-n + \frac{\omega - k_x v_x}{\Omega_\alpha(z)})\varphi']. \qquad (29)$$

$$\exp[-\frac{m_\alpha}{2\theta_\alpha}(\alpha_\alpha v_\perp^2 \cos^2\varphi' + 2\alpha_\alpha \frac{e_\alpha}{m_\alpha}A_y(z)v_\perp\cos\varphi')].$$

Expansion of this function under the integral in a Fourier series:

$$\exp[-\frac{m_\alpha}{2\theta_\alpha}(\alpha_\alpha v_\perp^2 \cos^2\varphi' + 2\alpha_\alpha \frac{e_\alpha}{m_\alpha}A_y(z)v_\perp\cos\varphi')] = \sum_{s=-\infty}^{\infty} C_s \exp(is\varphi'), \qquad (30)$$

Where:

$$C_s^{xx} = \frac{2}{\pi}\int_{\frac{\pi}{2}(\alpha-1)}^{\frac{\pi}{2}\alpha} \exp(-i4s\varphi')\exp[-\frac{m_\alpha}{2\theta_\alpha}(\alpha_\alpha v_\perp^2 \cos^2\varphi' + 2\alpha_\alpha \frac{e_\alpha}{m_\alpha}v_\perp A_y(z)\cos\varphi')]d\varphi' \qquad (31)$$

will allow obtaining from (29):

$$\varepsilon_{xx} = 1 + A\exp[-\frac{m_\alpha}{2\theta_\alpha}(v_x^2 + v_\perp^2)]\exp[-\frac{m_\alpha}{2\theta_\alpha}(\frac{\alpha_\alpha e_\alpha^2 A_y^2(z)}{m_\alpha^2})]\cdot$$

$$\sum_n J_n(b_\alpha)\sum_s C_s \int_\infty^\varphi d\varphi' \exp[i(k - n + \frac{\omega - k_x v_x}{\Omega_\alpha(z)})\varphi'].$$

Integration of the last expression will allow derivation of:

$$\varepsilon_{xx} = 1 + A\exp[-\frac{m_\alpha}{2\theta_\alpha}(v_x^2 + v_\perp^2)]\exp[-\frac{m_\alpha}{2\theta_\alpha}(\frac{\alpha_\alpha e_\alpha^2 A_y^2(z)}{m_\alpha^2})]\cdot$$

$$\sum_n J_n(b_\alpha)\sum_s C_s \frac{\exp[i(k - n + \frac{\omega - k_x v_x}{\Omega_\alpha(z)})\varphi]}{i(k - n + \frac{\omega - k_x v_x}{\Omega_\alpha(z)})}. \qquad (32)$$

At integration it has been considered that lower limit of integration $\varphi = \infty$ value of the reconstructed function goes to zero as the frequency has small imaginary positive part $\omega \to \omega + i\delta$ (the adiabatic hypothesis: perturbation of distribution function $\delta f_\alpha$ must disappear at $t \to -\infty$; in the assumed function of time $\delta f_\alpha \approx \exp(-i\omega t)$ such disappearance means that frequency has at least infinitely small positive imaginary part).

We will perform the integration on velocities in cylindrical coordinate system $d\vec{v} = dv_x v_\perp dv_\perp d\varphi$. Substituting the value of coefficient A (27), using decomposition (28) and making necessary transformations will allow obtaining from (32):

$$\varepsilon_{xx} = 1 - \sum_\alpha \frac{e_\alpha^2}{\varepsilon_0 \omega \theta_\alpha} \left(\frac{m_\alpha}{2\pi\theta_\alpha}\right)^{\frac{3}{2}} n_{0\alpha}(1+\alpha_\varepsilon)^{\frac{1}{2}} \exp[-\frac{m_\alpha}{2\theta_\alpha}(\frac{\alpha_\alpha e_\alpha^2 A_y^2(z)}{m_\alpha^2})] \cdot$$

$$\int dv_x v_\perp dv_\perp v_x^2 \exp[-\frac{m_\alpha}{2\theta_\alpha}(v_x^2+v_\perp^2)] \sum_n J_n(b_\alpha) \sum_s C_s \frac{1}{[\omega - k_x v_x - (n-k)\Omega_\varepsilon(z)]}. \quad (33)$$

$$\sum_l J_l(b_\alpha) \int_0^{2\pi} d\varphi \exp[il\varphi]\exp[-i(n-k)\varphi].$$

In this expression

$$\int_0^{2\pi} d\varphi \exp[il\varphi]\exp[-i(n-k)\varphi] = 2\pi, \text{ if } n\text{-}k = l,$$

$$0, \text{ ifи } n\text{-}k \neq l.$$

Then relation (33) becomes:

$$\varepsilon_{xx} = 1 - \sum_\alpha \frac{2\pi e_\alpha^2}{\varepsilon_0 \omega \theta_\alpha} \left(\frac{m_\alpha}{2\pi\theta_\alpha}\right)^{\frac{3}{2}} n_{0\alpha}(1+\alpha_\varepsilon)^{\frac{1}{2}} \exp[-\frac{m_\alpha}{2\theta_\alpha}(\frac{\alpha_\alpha e_\alpha^2 A_y^2(z)}{m_\alpha^2})] \sum_l \sum_s$$

$$\int_0^\infty C_s v_\perp \exp(-\frac{m_\alpha v_\perp^2}{2\theta_\alpha}) J_{k+l}(\frac{k_\perp v_\perp}{\Omega_\alpha(z)}) J_l(\frac{k_\perp v_\perp}{\Omega_\alpha(z)}) dv_\perp \int_{-\infty}^{+\infty} \frac{v_x^2 \exp(-\frac{m_\alpha v_x^2}{2\theta_\alpha})}{(\omega - k_x v_x - l\Omega_\alpha(z))} dv_x. \quad (34)$$

Finally, the component of tensor of dielectric permeability $\varepsilon_{xx}$ looks like:

$$\varepsilon_{xx} = 1 - \sum_\alpha B_\alpha \sum_s \sum_l G_{xx}(s,l,m_\alpha,\theta_\alpha,k_\perp,\Omega_\alpha(z)) I_{xx}(\beta). \quad (35)$$

Here,

$$B_\alpha = \frac{2\pi e_\alpha^2}{\varepsilon_0 \omega}(\frac{m_\alpha}{2\pi\theta_\alpha})^{\frac{3}{2}} \frac{n_{0\alpha}}{\theta_\alpha}(1+\alpha_\alpha)^{\frac{1}{2}} \exp[-\frac{m_\alpha}{2\theta_\alpha}(\frac{\alpha_\alpha e_\alpha^2 A_y^2(z)}{m_\alpha^2})], \quad (36)$$

$$C_s^{xx} = \frac{2}{\pi} \int_{\frac{\pi}{2}(\alpha-1)}^{\frac{\pi}{2}\alpha} \exp(-i4s\varphi')\exp[-\frac{m_\alpha}{2\theta_\alpha}(\alpha_\alpha v_\perp^2 \cos^2\varphi' + 2\alpha_\alpha \frac{e_\alpha}{m_\alpha} v_\perp A_y(z)\cos\varphi')]d\varphi' \quad (37)$$

$$G_{xx}(s,l,m_\alpha,\theta_\alpha,k_\perp,\Omega_\alpha(z)) = \int_0^\infty C_s v_\perp \exp(-\frac{m_\alpha v_\perp^2}{2\theta_\alpha}) J_{k+l}(\frac{k_\perp v_\perp}{\Omega_\alpha(z)}) J_l(\frac{k_\perp v_\perp}{\Omega_\alpha(z)}) dv_\perp, \quad (38)$$

$$I_{xx}(\beta) = \int_{-\infty}^{+\infty} \frac{v_x^2 \exp(-\frac{m_\alpha v_x^2}{2\theta_\alpha})}{(\omega - k_x v_x - l\Omega_\alpha(z))} dv_x = \frac{\theta_\alpha}{m_\alpha k_x}[\frac{\sqrt{2\pi}}{\beta} - i\pi\beta^2 \exp(-\frac{\beta^2}{2})], \text{если } \beta >> 1; \quad (39)$$

The remaining 8 components of tensor of dielectric permeability $\varepsilon_{ij}$ are calculated in a similar way. They are shown in the Application.

### 4.2. Derivation of dispersing equation.

Study of amplitude-frequency characteristics of perturbations of a transition layer between plasma and a magnetic field is based on the solution of Maxwell's equations for the perturbations of electromagnetic field closed by the constitutive equation. After substituting decompositions of electromagnetic field (3) in Maxwell'd equations (2), we will obtain the following combined equations:

$$\begin{cases} c^2 \varepsilon_0 rot\delta\vec{B} = \dfrac{\partial \delta\vec{D}}{\partial t}, div\delta\vec{B} = 0, \\ rot\delta\vec{E} = -\dfrac{\partial \delta\vec{B}}{\partial t}, div\delta\vec{D} = 0, \end{cases} \quad (40)$$

Where, $\delta\vec{B}(\vec{r},z,t)$ and $\delta\vec{E}(\vec{r},z,t)$ - perturbations of induction density and dielectric field intensity of the system under consideration;
$\delta D(\vec{r},z,t)$ - perturbation of dielectric density.

Maxwell's equations are completed here with the constitutive equation

$$\delta D_i(t,\vec{r},z) = \varepsilon_0 \int_{-\infty}^{t} dt' \int d\vec{r}' \mathcal{E}_{ij}(t-t',\vec{r}-\vec{r}',z) \delta E_j(t',\vec{r}',z). \quad (41)$$

Substituting this equation of expansions in plane waves (9), we will obtain a constitutive equation for Fourier amplitudes:

$$\delta D_i = \varepsilon_0 \varepsilon_{ij}(\omega,\vec{k},z) \delta E_j. \quad (42)$$

Here,

$$\varepsilon_{ij}(\omega,\vec{k},z) = \int_0^\infty dt_1 \int d\vec{r}_1 \mathcal{E}_{ij}(t_1,\vec{r}_1,z) \exp(i\omega t_1 - i\vec{k}\vec{r}_1) \quad (43)$$

- is a determination of tensor of dielectric permeability, $t_1 = t - t', \vec{r}_1 = \vec{r} - \vec{r}'$. This tensor was calculated above.

After substituting the expansions in plane waves (15), we will obtain from Maxwell's equations (40) the equation for perturbation of electric field $\delta E$:

$$[k^2 \delta_{ij} - k_i k_j - \dfrac{\omega^2}{c^2} \varepsilon_{ij}(\omega,\vec{k})] \delta E_j = 0. \quad (44)$$

A prerequisite to nontrivial solution of this system is that its determinant is equal to zero:

$$k^2 \delta_{ij} - k_i k_j - \dfrac{\omega^2}{c^2} \varepsilon_{ij}(\omega,\vec{k}) = 0. \quad (45)$$

The last relation is a dispersing equation for boundary layer perturbations.

## 5. Заключение

A procedure that allows study of unstable stability of a boundary layer between plasma and a magnetic field has been developed. Layer equilibrium for one reason or another is not set but follows from strict solution of kinetic equation with self-consistent electromagnetic field. Nicolson's solution was used as equilibrium solution. The non-equilibrium component to the distribution function was found and on this basis tensor of dielectric permeability simulating the medium was (boundary layer) is calculated. This tensor essentially differs from the known tensor for magnetoactive plasma. Equations for electric field perturbation were obtained from the Maxwell's equations completed with the derived constitutive equation. The technique of determination of dispersion characteristics of perturbations of a boundary layer was proposed. The theory is applicable to thermal nonrelativistic plasma.

**Application**

Throughout the paper:

$$B_\alpha = \frac{2\pi e_\alpha^2}{\varepsilon_0 \omega} \left(\frac{m_\alpha}{2\pi\theta_\alpha}\right)^{\frac{3}{2}} \frac{n_{0\alpha}}{\theta_\alpha} (1+\alpha_\alpha)^{\frac{1}{2}} \exp\left[-\frac{m_\alpha}{2\theta_\alpha}\left(\frac{\alpha_\alpha e_\alpha^2 A_y^2(z)}{m_\alpha^2}\right)\right], \tag{36}$$

$$C_s^{xx} = \frac{2}{\pi} \int_{\frac{\pi}{2}(\alpha-1)}^{\frac{\pi}{2}\alpha} \exp(-i4s\varphi')\exp\left[-\frac{m_\alpha}{2\theta_\alpha}(\alpha_\alpha v_\perp^2 \cos^2\varphi' + 2\alpha_\alpha \frac{e_\alpha}{m_\alpha} v_\perp A_y(z)\cos\varphi')\right]d\varphi' \tag{37}$$

## TENSOR COMPONENTS

$$\varepsilon_{xx} = 1 - \sum_\alpha B_\alpha \sum_s \sum_l G_{xx}(s,l,m_\alpha,\theta_\alpha,k_\perp,\Omega_\alpha(z))I_{xx}(\beta). \tag{35}$$

Here

$$G_{xx}(s,l,m_\alpha,\theta_\alpha,k_\perp,\Omega_\alpha(z)) = \int_0^\infty C_s^{xx} v_\perp \exp\left(-\frac{m_\alpha v_\perp^2}{2\theta_\alpha}\right) J_{s+l}\left(\frac{k_\perp v_\perp}{\Omega_\alpha(z)}\right) J_l\left(\frac{k_\perp v_\perp}{\Omega_\alpha(z)}\right) dv_\perp, \tag{38}$$

$$I_{xx}(\beta) = \int_{-\infty}^{+\infty} \frac{v_x^2 \exp(-\frac{m_\alpha v_x^2}{2\theta_\alpha})}{(\omega - k_x v_x - l\Omega_\alpha(z))} dv_x = \frac{\theta_\alpha}{m_\alpha k_x}\left[\frac{\sqrt{2\pi}}{\beta} - i\pi\beta^2 \exp\left(-\frac{\beta^2}{2}\right)\right], \text{если } \beta \gg 1, \tag{39}$$

$$C_s^{xx} = \frac{2}{\pi} \int_{\frac{\pi}{2}(\alpha-1)}^{\frac{\pi}{2}\alpha} \exp(-i4s\varphi')\exp\left[-\frac{m_\alpha}{2\theta_\alpha}(\alpha_\alpha v_\perp^2 \cos^2\varphi' + 2\alpha_\alpha \frac{e_\alpha}{m_\alpha} v_\perp A_y(z)\cos\varphi')\right]d\varphi'. \tag{37}$$

$$\varepsilon_{xy} = -\sum_\alpha \frac{\alpha_\alpha e_\alpha A_y(z)}{2m_\alpha} B_\alpha \sum_s \sum_l G_{xy}(s,l,m_\alpha,\theta_\alpha,k_\perp,\Omega_\alpha(z))I_{xy}(\beta). \tag{A1}$$

Here

$$G_{xy}(s,l,m_\alpha,\theta_\alpha,k_\perp,\Omega_\alpha(z)) = \int_0^\infty C_s^{xy} v_\perp \exp\left(-\frac{m_\alpha v_\perp^2}{2\theta_\alpha}\right) J_{s+l}\left(\frac{k_\perp v_\perp}{\Omega_\alpha(z)}\right) J_l\left(\frac{k_\perp v_\perp}{\Omega_\alpha(z)}\right) dv_\perp, \tag{A2}$$

$$I_{xy}(\beta) = \int_{-\infty}^{+\infty} \frac{v_x \exp(-\frac{m_\alpha v_x^2}{2\theta_\alpha})}{(\omega - k_x v_x - l\Omega_\alpha(z))} dv_x \approx \frac{1}{k_x}\sqrt{\frac{\theta_\alpha}{m_\alpha}} \frac{\sqrt{2\pi}}{\beta^2}, \quad \beta \gg 1, \tag{A3}$$

$$C_s^{xy} = \frac{2}{\pi} \int_{\frac{\pi}{2}(\alpha-1)}^{\frac{\pi}{2}\alpha} \exp(-i4s\varphi') \cdot [\frac{(2+\alpha_\alpha)m_\alpha}{2\theta_\alpha} v_\perp \cos\varphi' + \frac{\alpha_\alpha e_\alpha}{2\theta_\alpha} A_y(z)] \cdot \exp[-\frac{m_\alpha}{2\theta_\alpha}(\alpha_\alpha v_\perp^2 \cos^2\varphi' +$$

$$+ 2\alpha_\alpha \frac{e_\alpha}{m_\alpha} v_\perp A_y(z) \cos\varphi')] d\varphi'.$$

(A4)

$$\varepsilon_{xz} = -\sum_\alpha B_\alpha \sum_s \sum_l G_{xz}(s,l,m_\alpha,\theta_\alpha,k_\perp,\Omega_\alpha(z)) I_{xz}(\beta).$$

(A5)

Here

$$G_{xz}(s,l,m_\alpha,\theta_\alpha,k_\perp,\Omega_\alpha(z)) = \int_0^\infty C_s^{xz} v_\perp^2 \exp(-\frac{m_\alpha v_\perp^2}{2\theta_\alpha}) J_{s+l}(\frac{k_\perp v_\perp}{\Omega_\alpha(z)}) J_l(\frac{k_\perp v_\perp}{\Omega_\alpha(z)}) dv_\perp$$

, (A6)

$$I_{xz}(\beta) = \int_{-\infty}^{+\infty} \frac{v_x \exp(-\frac{m_\alpha v_x^2}{2\theta_\alpha})}{(\omega - k_x v_x - l\Omega_\alpha(z))} dv_x \approx \frac{1}{k_x} \sqrt{\frac{\theta_\alpha}{m_\alpha}} \frac{\sqrt{2\pi}}{\beta^2}, \beta \gg 1,$$

(A7)

$$C_s^{xz} = \frac{2}{\pi} \int_{\frac{\pi}{2}(\alpha-1)}^{\frac{\pi}{2}\alpha} \exp(-i4s\varphi') \cdot \sin\varphi' \cdot \exp[-\frac{m_\alpha}{2\theta_\alpha}(\alpha_\alpha v_\perp^2 \cos^2\varphi' + 2\alpha_\alpha \frac{e_\alpha}{m_\alpha} v_\perp A_y(z) \cos\varphi')] d\varphi'$$

(A8)

$$\varepsilon_{yx} = -\sum_\alpha B_\alpha \sum_s \sum_l G_{yx}(s,l,m_\alpha,\theta_\alpha,k_\perp,\Omega_\alpha(z)) I_{yx}(\beta).$$ (A9)

Here

$$G_{yx}(s,l,m_\alpha,\theta_\alpha,k_\perp,\Omega_\alpha(z)) = \int_0^\infty C_s^{yx} v_\perp^2 \exp(-\frac{m_\alpha v_\perp^2}{2\theta_\alpha}) J_{s+l}(\frac{k_\perp v_\perp}{\Omega_\alpha(z)}) l J_l(\frac{k_\perp v_\perp}{\Omega_\alpha(z)}) \frac{\Omega_\alpha(z)}{k_\perp v_\perp} dv_\perp,$$ (A10)

$$I_{yx}(\beta) = \int_{-\infty}^{+\infty} \frac{v_x \exp(-\frac{m_\alpha v_x^2}{2\theta_\alpha})}{(\omega - k_x v_x - l\Omega_\alpha(z))} dv_x \approx \frac{1}{k_x} \sqrt{\frac{\theta_\alpha}{m_\alpha}} \frac{\sqrt{2\pi}}{\beta^2}, \beta \gg 1,$$ (A11)

$$C_s^{yx} = \frac{2}{\pi} \int_{\frac{\pi}{2}(\alpha-1)}^{\frac{\pi}{2}\alpha} \exp(-i4s\varphi') \cdot \exp[-\frac{m_\alpha}{2\theta_\alpha}(\alpha_\alpha v_\perp^2 \cos^2\varphi' + 2\alpha_\alpha \frac{e_\alpha}{m_\alpha} v_\perp A_y(z) \cos\varphi')] d\varphi'.$$ (A12)

$$\varepsilon_{yy} = 1 - \sum_\alpha \frac{\alpha_\alpha e_\alpha A_y(z)}{2m_\alpha} B_\alpha \sum_s \sum_l G_{yy}(s,l,m_\alpha,\theta_\alpha,k_\perp,\Omega_\alpha(z)) I_{yy}(\beta).$$ (A13)

Here

$$G_{yy}(s,l,m_\alpha,\theta_\alpha,k_\perp,\Omega_\alpha(z)) = \int_0^\infty C_s^{yy} v_\perp^2 \exp(-\frac{m_\alpha v_\perp^2}{2\theta_\alpha}) J_{s+l}(\frac{k_\perp v_\perp}{\Omega_\alpha(z)}) l J_l(\frac{k_\perp v_\perp}{\Omega_\alpha(z)}) \frac{\Omega_\alpha(z)}{k_\perp v_\perp} dv_\perp \ , \quad (A14)$$

$$I_{yy}(\beta) = \int_{-\infty}^{+\infty} \frac{\exp(-\frac{m_\alpha v_x^2}{2\theta_\alpha})}{(\omega - k_x v_x - l\Omega_\alpha(z))} dv_x \approx \frac{1}{k_x} \frac{\sqrt{2\pi}}{\beta} \ , \ \beta \gg 1, \quad (A15)$$

$$C_s^{yy} = \frac{2}{\pi} \int_{\frac{\pi}{2}(\alpha-1)}^{\frac{\pi}{2}\alpha} \exp(-i4s\varphi') \cdot [\frac{(2+\alpha_\alpha)m_\alpha}{2\theta_\alpha} v_\perp \cos\varphi' + \frac{\alpha_\alpha e_\alpha}{2\theta_\alpha} A_y(z)] \cdot \exp[-\frac{m_\alpha}{2\theta_\alpha}(\alpha_\alpha v_\perp^2 \cos^2\varphi' + \quad (A16)$$

$$+ 2\alpha_\alpha \frac{e_\alpha}{m_\alpha} v_\perp A_y(z) \cos\varphi')] d\varphi'.$$

$$\varepsilon_{yz} = -\sum_\alpha B_\alpha \sum_s \sum_l G_{yz}(s,l,m_\alpha,\theta_\alpha,k_\perp,\Omega_\alpha(z)) \cdot I_{yz}(\beta). \quad (A17)$$

Here

$$G_{yz}(s,l,m_\alpha,\theta_\alpha,k_\perp,\Omega_\alpha(z)) = \int_0^\infty C_s^{yz} v_\perp^3 \exp(-\frac{m_\alpha v_\perp^2}{2\theta_\alpha}) J_{s+l}(\frac{k_\perp v_\perp}{\Omega_\alpha(z)}) l J_l(\frac{k_\perp v_\perp}{\Omega_\alpha(z)}) \frac{\Omega_\alpha(z)}{k_\perp v_\perp} dv_\perp \ , \quad (A18)$$

$$I_{yz}(\beta) = \int_{-\infty}^{+\infty} \frac{\exp(-\frac{m_\alpha v_x^2}{2\theta_\alpha})}{(\omega - k_x v_x - l\Omega_\alpha(z))} dv_x \approx \frac{1}{k_x} \frac{\sqrt{2\pi}}{\beta} \ , \ \beta \gg 1, \quad (A19)$$

$$C_s^{yz} = \frac{2}{\pi} \int_{\frac{\pi}{2}(\alpha-1)}^{\frac{\pi}{2}\alpha} \exp(-i4s\varphi') \cdot \sin\varphi' \cdot \exp[-\frac{m_\alpha}{2\theta_\alpha}(\alpha_\alpha v_\perp^2 \cos^2\varphi' + 2\alpha_\alpha \frac{e_\alpha}{m_\alpha} v_\perp A_y(z) \cos\varphi')] d\varphi'. \quad (A20)$$

$$\varepsilon_{zx} = i\sum_\alpha B_\alpha \sum_s \sum_l G_{zx}(s,l,m_\alpha,\theta_\alpha,k_\perp,\Omega_\alpha(z)) \cdot I_{zx}(\beta). \quad (A21)$$

Here

$$G_{zx}(s,l,m_\alpha,\theta_\alpha,k_\perp,\Omega_\alpha(z)) = \int_0^\infty C_s^{zx} v_\perp^2 \exp(-\frac{m_\alpha v_\perp^2}{2\theta_\alpha}) J_{s+l}(\frac{k_\perp v_\perp}{\Omega_\alpha(z)}) J_l'(\frac{k_\perp v_\perp}{\Omega_\alpha(z)}) dv_\perp \ , \quad (A22)$$

where:

$$J_l'(\frac{k_\perp v_\perp}{\Omega_\alpha(z)}) = \frac{1}{2}(J_{l-1}(\frac{k_\perp v_\perp}{\Omega_\alpha(z)}) - J_{l+1}(\frac{k_\perp v_\perp}{\Omega_\alpha(z)})).$$

$$I_{zx}(\beta) = \int_{-\infty}^{+\infty} \frac{v_x \exp(-\frac{m_\alpha v_x^2}{2\theta_\alpha})}{(\omega - k_x v_x - l\Omega_\alpha(z))} dv_x \approx \frac{1}{k_x} \sqrt{\frac{\theta_\alpha}{m_\alpha}} \frac{\sqrt{2\pi}}{\beta^2} \ , \ \beta \gg 1, \quad (A23)$$

$$C_s^{zx} = \frac{2}{\pi} \int_{\frac{\pi}{2}(\alpha-1)}^{\frac{\pi}{2}\alpha} \exp(-i4s\varphi') \exp[-\frac{m_\alpha}{2\theta_\alpha}(\alpha_\alpha v_\perp^2 \cos^2\varphi' + 2\alpha_\alpha \frac{e_\alpha}{m_\alpha} v_\perp A_y(z)\cos\varphi')]d\varphi'. \tag{A24}$$

$$\varepsilon_{zy} = i\sum_\alpha \frac{\alpha_\alpha e_\alpha A_y(z)}{2m_\alpha} B_\alpha \sum_s \sum_l G_{zy}(s,l,m_\alpha,\theta_\alpha,k_\perp,\Omega_\alpha(z)) \cdot I_{zy}(\beta). \tag{A25}$$

Where

$$G_{zy}(s,l,m_\alpha,\theta_\alpha,k_\perp,\Omega_\alpha(z)) = \int_0^\infty C_s^{zy} v_\perp^2 \exp(-\frac{m_\alpha v_\perp^2}{2\theta_\alpha}) J_{s+l}(\frac{k_\perp v_\perp}{\Omega_\alpha(z)}) J_l'(\frac{k_\perp v_\perp}{\Omega_\alpha(z)}) dv_\perp, \tag{A26}$$

where:

$$J_l'(\frac{k_\perp v_\perp}{\Omega_\alpha(z)}) = \frac{1}{2}(J_{l-1}(\frac{k_\perp v_\perp}{\Omega_\alpha(z)}) - J_{l+1}(\frac{k_\perp v_\perp}{\Omega_\alpha(z)})).$$

$$I_{zy}(\beta) = \int_{-\infty}^{+\infty} \frac{\exp(-\frac{m_\alpha v_x^2}{2\theta_\alpha})}{(\omega - k_x v_x - l\Omega_\alpha(z))} dv_x \approx \frac{1}{k_x}\frac{\sqrt{2\pi}}{\beta}, \ \beta \gg 1, \tag{A27}$$

$$C_s^{zy} = \frac{2}{\pi} \int_{\frac{\pi}{2}(\alpha-1)}^{\frac{\pi}{2}\alpha} \exp(-i4s\varphi') \cdot [\frac{(2+\alpha_\alpha)m_\alpha}{2\theta_\alpha} v_\perp \cos\varphi' + \frac{\alpha_\alpha e_\alpha}{2\theta_\alpha} A_y(z)] \cdot \exp[-\frac{m_\alpha}{2\theta_\alpha}(\alpha_\alpha v_\perp^2 \cos^2\varphi' +$$
$$+ 2\alpha_\alpha \frac{e_\alpha}{m_\alpha} v_\perp A_y(z)\cos\varphi')]d\varphi'. \tag{A28}$$

$$\varepsilon_{zz} = 1 + i\sum_\alpha B_\alpha \sum_s \sum_l G_{zz}(s,l,m_\alpha,\theta_\alpha,k_\perp,\Omega_\alpha(z)) \cdot I_{zz}(\beta). \tag{A29}$$

Here

$$G_{zz}(s,l,m_\alpha,\theta_\alpha,k_\perp,\Omega_\alpha(z)) = \int_0^\infty C_s^{zz} v_\perp^3 \exp(-\frac{m_\alpha v_\perp^2}{2\theta_\alpha}) J_{s+l}(\frac{k_\perp v_\perp}{\Omega_\alpha(z)}) J_l'(\frac{k_\perp v_\perp}{\Omega_\alpha(z)}) dv_\perp, \tag{A30}$$

where:

$$J_l'(\frac{k_\perp v_\perp}{\Omega_\alpha(z)}) = \frac{1}{2}(J_{l-1}(\frac{k_\perp v_\perp}{\Omega_\alpha(z)}) - J_{l+1}(\frac{k_\perp v_\perp}{\Omega_\alpha(z)})).$$

$$I_{zz}(\beta) = \int_{-\infty}^{+\infty} \frac{\exp(-\frac{m_\alpha v_x^2}{2\theta_\alpha})}{(\omega - k_x v_x - l\Omega_\alpha(z))} dv_x \approx \frac{1}{k_x}\frac{\sqrt{2\pi}}{\beta}, \ \beta \gg 1, \tag{A31}$$

$$C_s^{zz} = \frac{2}{\pi} \int_{\frac{\pi}{2}(\alpha-1)}^{\frac{\pi}{2}\alpha} \exp(-i4s\varphi') \cdot \sin\varphi' \cdot \exp[-\frac{m_\alpha}{2\theta_\alpha}(\alpha_\alpha v_\perp^2 \cos^2\varphi' + 2\alpha_\alpha \frac{e_\alpha}{m_\alpha} v_\perp A_y(z)\cos\varphi')]d\varphi'. \tag{A32}$$